\documentstyle[aps,prb,graphicx]{revtex}
\begin{document}
\draft
\twocolumn[\hsize\textwidth\columnwidth\hsize\csname 
@twocolumnfalse\endcsname
\title{Bound states in $d$-density-wave phases}
\author{Carsten Honerkamp and Manfred Sigrist}

\address{Theoretische Physik,
ETH-H\"onggerberg, CH-8093 Z\"urich, Switzerland }
\date{July 26, 2001}
\maketitle 
\begin{abstract} 
We investigate the quasiparticle spectrum near surfaces in a
two-dimensional system with $d$-density-wave order within a 
mean-field theory.  
For Fermi surfaces with perfect nesting for the
ordering wave vector $(\pi,\pi)$ of the $d$-density-wave, 
a zero energy bound state 
occurs at 110 surfaces, in close analogy with the known effect in
$d$-wave superconducting states or graphite. When the shape of the Fermi
surface is changed by doping, the bound 
state energy moves away from the Fermi level. Furthermore, away from
half-filling we find inhomogeneous phases with domain walls of
the $d$-density-wave order parameter. 
The domain walls also support low energy bound states. 
These phenomena might provide an experimental test for hidden $d$-density-wave
 order in the high-$T_c$ cuprates.

\end{abstract}
\vskip1pc]
\narrowtext
\section{Introduction}
In the search for an explanation of the so-called pseudogap regime of the
underdoped high-$T_c$ cuprates, Chakravarty, Laughlin, Morr, and
Nayak\cite{chakravarty} 
recently proposed a novel type of order, known as $d$-density-wave order. 
This state was discussed earlier by Schulz\cite{schulz} and in more detail by
Nayak\cite{nayak}. It breaks translational and time-reversal symmetry
and is related 
to flux phases known from 
$t$-$J$ and Hubbard models at infinite or large onsite
repulsion\cite{affleck,lee}.  
Its order parameter, written with the electrons' creation and
annihilation operators and ordering wavevector $\vec{Q}=(\pi,\pi)$, 
\begin{equation}
\Phi_{\mathrm{ddw}} = \sum_{\vec{k},s} f(\vec{k}) \, \langle
c^\dagger_{\vec{k},s} 
c_{\vec{k}+\vec{Q},s} \rangle \label{ddwop}
\end{equation}
is purely imaginary due to the sign change of the $d_{x^2-y^2}$-wave form
factor
\begin{equation}
f(\vec{k}) \sim \cos k_x - \cos k_y \label{dwff} 
\end{equation}
under $\vec{k} \to \vec{k}+\vec{Q}$. In real space  it gives rise to
circulating in-plane currents around elementary plaquettes. The 
orientation of the current flow, i.e. the sign of the flux penetrating the
plaquettes, is alternating from plaquette to plaquette. 

 Since the staggered flux order parameter 
is of Ising type the $d$-density-wave can
order in two spatial dimensions at nonzero temperatures. In one 
spatial dimension it can still lead to long range order in
the ground state, as recently shown for the two-leg ladder by
 Fjaerestad and Marston\cite{marston}.

Apart from the implications of the proposed type of order on observable bulk
quantities like specific heat, spin susceptibilities and conductivities
\cite{chakravarty,dhlee,csting}, the $d$-wave structure of the order
parameter opens 
another field of phenomena. These are surface effects, which have
already played a 
significant role in the experimental attempts to
identify the {\em superconducting} pairing symmetries
in the high-$T_c$ cuprates and other unconventional
superconductors\cite{vanharlingen,sigrist}. 
Explicitly we mention the possibility of Andreev bound
states at surfaces of unconventional superconductors
\cite{unctunnel,greene,alff,deutscher,yeh,waelti,laube} with
 suitable orientation (and quality in experimental
studies) at low or zero energies. The theoretically appealing feature
of these surface bound 
states is that they are a consequence of the symmetry of the order
parameter and 
depend less on the details of the (most of the time drastically simplified)
models used to calculate them.
Since the $d$-density-wave order described above also occurs with a non-trivial
symmetry, it appears to be interesting to search for surface effects
in a $d$-density 
wave ordered state.  Related to latter phenomena are of course local
deformations of the bulk state at impurities. 
In the context with $d$-density-wave states, these were recently studied
theoretically by
Zhu et al.\cite{zhu}, Wang\cite{wang}, and Morr\cite{morr}, 
which all found low energy states in the vicinity of non-magnetic impurities.

In this paper we analyze the possibility of surface bound states for 
electrons on a two-dimensional square lattice with nearest neighbor hopping in 
the presence of $d$-density-wave order. 
After a simplified analytical treatment 
showing the existence of surface bound states at [110] surfaces, 
we solve the mean-field equations for the $d$-density-wave amplitude 
self-consistently on a square lattice strip with finite width.  The
$d$-density-wave order is generated by a repulsive nearest neighbor
density-density interaction. 
We find low energy surface states at the [110] surface of the lattice
and study their 
dependence on the band filling. Further we analyze the $d$-density-wave state
away from perfect $(\pi,\pi)$-nesting at half filling and find 
indications for domain walls of the order parameter forming between locally 
half-filled $d$-density-wave domains. These domain walls themselves
can again host 
bound states with
sub-gap energies.

Of course a realistic description of the high-$T_c$ cuprates may
require a more sophisticated 
theory, so that our simplified mean-field treatment might neglect a 
substantial part of the physics. Nevertheless 
we emphasize that the occurrence of surface bound states relies 
mainly on the symmetry of the order parameter. Thus our findings
should be a robust feature
remaining true in more detailed studies. Moreover the theoretical analysis of 
a large number of experiments in the framework of $d$-density-wave order by
Chakravarty et al.\cite{chakravarty} was performed on a comparable
level of approximations.

\section{Analytical treatment}
In order to understand the main features at the surface, let us
consider electrons 
on a two-dimensional square lattice with hopping between nearest neighbors
at half filling.  
Then the free Fermi surface is the perfectly nested 
square.
We assume that the system has $d$-density-wave order, giving 
rise to a term in the Hamiltonian,
\begin{equation} H_{\mathrm{ddw}} = \sum_{\vec{k}} i W_{0}\,  (\cos
  k_{x}-\cos k_{y}) \left[  
c^\dagger_{\vec{k}} c_{\vec{k}+\vec{Q}} - c^\dagger_{\vec{k}+\vec{Q}} 
c_{\vec{k}} \right] . \label{ddwham} \end{equation}
after mean-field decoupling, 
with real $W_0$ and $\vec{Q}=(\pi,\pi)$ (in units of the inverse
lattice constant $a^{-1}$). 
In real space this term also couples nearest neighbor sites,
analogous to the ordinary hopping term, so that
the total Hamiltonian has 
bipartite structure. In the following we will use this property and
rewrite the Hamiltonian 
by doubling the unit cell and 
separating the wave function into components belonging to the $A$-
and $B$-sublattice, $ \Psi_A 
(\vec{r}) $ and $ \Psi_B (\vec{r}) $,

\begin{equation}
\left[ \begin{array}{cc} 
0 & \varepsilon_{\vec{k}} + i \gamma_{\vec{k}} \\ 
\varepsilon_{\vec{k}} - i \gamma_{\vec{k}} & 0 \end{array} \right]
\left( \begin{array}{c} \Psi_{A} (\vec{r}) \\ \Psi_{B} (\vec{r})
  \end{array} \right)  
= E \left( \begin{array}{c} \Psi_{A} (\vec{r}) \\ \Psi_{B} (\vec{r})
 \end{array} 
 \right)
\end{equation}
where $ \varepsilon_{\vec{k}} = -2t (\cos k_x + \cos k_y) $ and $
\gamma_{\vec{k}} = W_0  
(\cos k_x - \cos k_y ) $ and $ \vec{k} = -i \vec{\nabla} $. Obviously
this formulation leads to the proper excitation spectrum in the uniform case
which has two Fermi points on the boundary of the reduced Brillouin zone at 
$ \vec{k} = \vec{K} = (\frac{\pi}{2} , 
\frac{\pi}{2} ) $ and $ \vec{K}'= (\frac{\pi}{2}, - \frac{\pi}{2} )
$. Close to these points the  
spectrum has the approximate form of a Weyl spectrum for a chiral
particle with opposite chirality at $ \vec{K} $ and $ \vec{K}' $. 
In this respect the spectrum of the $d$-density-wave state 
has obvious similarity with the spectrum of a graphite sheet \cite{graphite}. 

Now we consider the problem of a surface with normal vector parallel
to the [110]-direction which 
defines the coordinate $ x' = (x+y)/\sqrt{2} $ ($ x' > 0 $ and the
surface is located at $ x'=0 $). Here we will
not attempt a self-consistent solution of the mean-field problem,
but keep the $d$-density-wave order parameter $ W_0 $ constant for
$ x' > 0 $. This is sufficient to show that this surface possess
 surface bound states at zero
energy. The translation symmetry along the direction parallel to the 
surface ($ y' = (x-y)/ \sqrt{2} $)
is maintained so that the momentum $ k_y' = (k_x - k_y)/\sqrt{2} $ 
is conserved. Let us consider
electronic states in the vicinity of the non-interacting Fermi
line $k_x'=(k_x + k_y)/\sqrt{2}= \pi$ (in units of inverse of the lattice
constant ${a^*}=\sqrt{2}a$ of the doubled unit cell) at the boundary of the reduced Brillouin
zone. We extract the
fast oscillating and $y$-dependent parts from the wave functions by writing
\begin{equation}
\Psi_{A(B)} (\vec{r}) = f_{A(B),k_y'} (x')\, e^{i \pi x'}
e^{i k_y' y'} \label{psiab}
\end{equation}
Next we linearize the kinetic energy $\varepsilon_{\vec{k}}$ around the
line $k_x'= \pi$ and introduce $p'_x = k_x'-\pi$. 
Then, for $ f_{A(B), k_y'} $ the Hamiltonian has the form
\begin{equation}
\left[ \begin{array}{cc} 0 & \epsilon_{\vec{k}}  \\
\epsilon_{\vec{k}}^* & 0 \end{array} \right] \left(
\begin{array}{c}  
f_{A, k_y'}(x') \\ f_{B, k_y'} (x') \end{array} \right) = E \left(
\begin{array}{c}  
f_{A, k_y'}(x') \\ f_{B, k_y'} (x') \end{array} \right)
\end{equation}
where \[ \epsilon_{\vec{k}} = v_{F,x'} (k_y') \, p'_x + 2 i W_0 \sin\left( \frac{
k_y'}{2} \right) \, . \]
Here we replace $ p'_x = - i \partial_{x'} $. For the
[110]-surface the outer most sites belong to one sublattice, say $A$. 
Then we assume that the wave function
takes nonzero values only on this sublattice and vanishes on the other. 
This behavior of the bound state wave function
 is confirmed by the self-consistent solution described below.
Thus, we impose the condition 
$ f_{B, k_y'} (x') = 0 $ and $ E = 0 $ and analyze the conditions to
obtain a zero energy bound state from the  
Hamiltonian above\cite{igami}. The resulting equation is

\begin{equation}
-i \partial_{x'} f_{A, k_y'}(x') =  2 i \, \frac{W_0}{v_{F,x}}\,  \sin \left( \frac{
k_y'}{2} \right) \,   f_{A, k_y'}(x')
\end{equation}
which has an exponentially decaying bound state 
solution $ f_{A, k_y'} (x') =
C e^{-\kappa x'} $. The existence condition is $ k_y' >0 $ with
$ \kappa = 2 W_0 \sin \left( k_y'/2 \right) /v_{F,x}$ and $W_0>0$.
As $ k_y' $ approaches 0 the bound state extends 
deeper and deeper into the bulk. 
Zero energy bound states only exist for
one sign of the momentum parallel to the surface, an indication of
broken time reversal  
symmetry of the bulk phase. Furthermore, the bound state has
non-bonding character 
as it occupies only one sublattice. Applying the time reversal operation
or interchanging $A$- and $B$-sublattices reverses $ k_y' $. 
A related situation occurs at so-called 
zigzag edges of graphite where the bound state spectrum has similar
properties, but does not violate time reversal symmetry \cite{igami,zigzag}. 

\begin{figure}
\begin{center} 
\includegraphics[width=.48\textwidth]{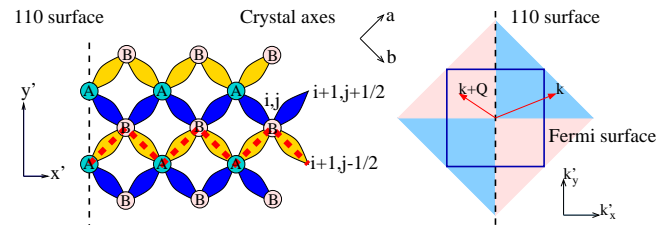}
\end{center} 
\caption{The [110] surface of the two-dimensional square lattice. Left: real
space picture. The
  circles denote the 
lattice sites of $A$- and $B$-sublattice, and the differently shaded bonds between them indicate the sign
structure of the $d_{x^2-y^2}$-density-wave amplitude $\langle 
c^\dagger_{i,j}c_{i+1,j \pm 1/2} \rangle$. Right: $\vec{k}$-space view.}
\label{geo}
\end{figure} 
We may also include a small shift of the chemical potential $ \mu $ ($
\ll t $). Within our approximative scheme this merely leads 
to a shift of the bound state energy $ E $ to $ -\mu$. A similar shift 
of the spectrum was also obtained by Zhu et al.\cite{zhu}
and Wang\cite{wang} for the impurity case. 
Below, we will find the same effect in the self-consistent treatment.
 
Concluding this analysis we observe that the [110]-surface of a
$d$-density-wave ordered state on a two-dimensional square lattice
should support low-lying surface states. We note that at [100] surfaces 
 no bound state can form according to a Bogoliubov-deGennes analysis. 
In the next section we confirm our qualitative results for the [110] surface 
with a self-consistent
mean-field calculation using tight binding model, which also
takes into account the suppression on the $d$-density-wave order parameter at
the odd parity surface.

\section{Model and self-consistent mean-field scheme}
As a model for the [110] surface, we consider a strip of a rotated
square lattice  
with infinite extension along the surface (see Fig.\ \ref{geo}). For
simplicity we 
 neglect the effects of 
the $d$-density-wave order on the spin degrees of freedom.  Then for each spin
orientation the kinetic
energy through hopping between neighboring sites gives rise to a term
\begin{equation} H_t = -t \sum_{(i,j)} \left[ c^\dagger_{i,j} c_{i+1,j+1/2} +
c^\dagger_{i,j} c_{i+1,j-1/2} + c.c. \right] \label{hkin} \end{equation}
The index $i$ labels the lattice sites along a zig-zag chain in $x$-direction,
while the indices $i+1,j\pm 1/2$ label the right upper or lower
neighbors of the site $i,j$.  
Next we introduce a repulsive nearest neighbor interaction
\begin{eqnarray} H_V &=& V \sum_{i,j}  \left[ c^\dagger_{i,j} c_{i,j}
    c^\dagger_{i+1,j+1/2} 
c_{i+1,j+1/2} \right. \nonumber \\ && \left. \qquad \quad +\,
c^\dagger_{i,j} c_{i,j} c^\dagger_{i+1,j-1/2} 
c_{i+1,j-1/2} \right] \, . \label{hv} \end{eqnarray}
We have already left out the interaction between densities of spin-up and
spin-down electrons on neighboring sites, as this does not contribute to the
$d$-density-wave decoupling. In real space, the static
$d$-density-wave amplitude 
between nearest neighbors $\vec{x}$ and $\vec{y}$ has the following
form\cite{nayak}: 
\begin{eqnarray} W (\vec{x},\vec{y})& =& \langle c^\dagger_{\vec{x},s}
c_{\vec{y},s'} \rangle =
\frac{i}{2} \, W_0 \, \delta_{s,s'}  \, e^{i \vec{Q} \vec{x}} \nonumber \\&& 
\left( \delta_{\vec{y},\vec{x}-\hat{a}}
+\delta_{\vec{y},\vec{x}+\hat{a}}
- \delta_{\vec{y},\vec{x}-\hat{b}} -\delta_{\vec{y},\vec{x}+\hat{b}}
\right)\, \label{ddwreal} \end{eqnarray}
where   $\hat{a}$ and $\hat{b}$ denote vectors along the crystal axes with
length of the lattice constant $a$. In our rotated system (see
Fig.\ \ref{geo}), the mean-field decoupling 
is performed by introducing the $j$- or $y$-independent imaginary mean-field
\begin{equation} W_{i,i+1} = \frac{V}{2} \left[ \langle
    c^\dagger_{i,j}c_{i+1,j+1/2} \rangle 
-\langle c^\dagger_{i,j}c_{i+1,j-1/2} \rangle \right] \, \label{mfdef}
\end{equation} 
into the interaction term (\ref{hv}). The minus-sign between the two
expectation values takes care of the sign change of the $d_{x^2-y^2}$-symmetric
order parameter under $\pi/2$-rotation of the lattice. Furthermore for the
$d_{x^2-y^2}$-density-wave state (\ref{ddwreal}) we expect 
 $W_{i,i+1}$ to alternate $\propto
(-1)^i$, but we do not enforce this behavior in the numerical mean-field
treatment.
Using this notation, the maximal gap magnitude in the homogeneous bulk
state is $2 |W_{i,i+1}|$. 

Next we exploit the translational invariance along
$y$-direction by introducing the partial Fourier transform
\[ c_{i,j} =\frac{1}{\sqrt{N_y}} \sum_k c_{i,k} \, e^{i k y_j} \, . \]
$N_y$ denotes the number of lattice sites in $y$-direction, which we choose
large enough for the numerical calculation. Since the distance in the
$y$-direction between the zig-zag chains along the $x$-direction
 is  $a^{*}=\sqrt{2}a$,
$k=k'_y$ has to range from $-\pi/a^*$ to $\pi/a^*$. In the
following we set $a^*=1$.
With this the Hamiltonian becomes
\begin{eqnarray}
H&=& -2t \sum_{i,k} \left[ c^\dagger_{i,k}c_{i+1,k} +c^\dagger_{i+1,k}c_{i,k}
\right] \cos\frac{k}{2} \nonumber \\
&& - \mu \sum_{i,k} c^\dagger_{i,k} c_{i,k}  \nonumber \\
&& - i \sum_{i,k} W_{i,i+1}  \left[ c^\dagger_{i,k}c_{i+1,k} +
c^\dagger_{i+1,k}c_{i,k} \right]\, \sin \frac{k}{2} \, .\label{mfham}
\end{eqnarray} 
This Hamiltonian can be easily diagonalized for each $k$ on the finite
strip. The 
self-consistent scheme is complemented with the gap equation
\begin{equation} W_{i,i+1} = -\frac{i}{N} \sum_{k} \langle c^\dagger_{ik}
c_{i+1,k} \rangle \, \sin \frac{k}{2} \, . \label{ddwgapeq} \end{equation}
From this equation and the eigenfunctions of (\ref{mfham}) we can calculate
the local $d$-density-wave amplitude $W_{i,i+1}$. Inserting this back into
(\ref{mfham}) we can repeat the process until self-consistency is reached.

\section{Numerical results}
\subsection{Results for perfect nesting: zero energy states}
\begin{figure}
\begin{center} 
\includegraphics[width=.49\textwidth]{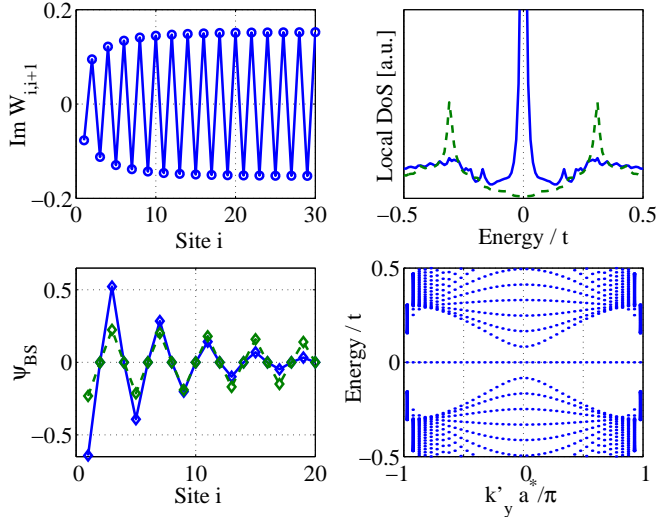}
\end{center} 
\caption{Upper left plot: Spatial variation of the (purely imaginary)
$d$-density-wave amplitude $W_{i,i+1}$. Upper right plot: Surface 
(on the first lattice site, solid line) 
and bulk (dashed line) density of states versus energy. Lower left plot:
Wave functions of two  
zero energy bound states corresponding to transverse wave vectors $k'_y \approx
-3\pi/4$ (solid line) and $-\pi /4$ (dashed line).  
The data shown are calculated for a system with 300 sites in
$x$-direction, $V=2t$, $\mu=0$ and $T=0.01t$. Lower right plot: Spectrum as
function of the transverse wavevector $k'_y$. The zero energy states for
positive and negative $k'_y$ are localized on opposite surfaces of the
strip. All states at nonzero energy extend over the whole system.} 
\label{mu0n300T10}
\end{figure}

Let us first discuss the results at half-filling. We choose $\mu=0$ and $V=2t$.
In the left plot Fig.\ \ref{mu0n300T10} we show the
spatial variation of the $d$-density amplitude at the surface. One clearly
observes the staggered real space structure of the mean-field $W_{i,i+1}$, in
agreement with (\ref{ddwreal}). At the surface the amplitude is
slightly suppressed on 
the length scale of a few lattice constants, similar to the suppression of the
$d$-wave superconducting gap function in analogous superconducting
cases\cite{unctunnel,honerkamp}.  
In the right plot we show the local density of states at the surface and in
the bulk of the system. The bulk density of states exhibits
a well formed gap edge at $2|W_{i,i+1}^{\mathrm{bulk}}|$ 
with some low energy states due to the nodes of the $d$-density-wave gap.
The surface density of states shows a pronounced zero energy peak. 
Studying the wave functions corresponding to these low energy
eigenvalues, we find  
that the peak indeed arises due to bound states (see lower plot of
Fig.\ \ref{mu0n300T10}), confirming the expectations from the
analysis of Sec.II. The bound state wave functions take nonzero values
only on one sublattice and their decay length depends on the
size of the $d$-density-wave bulk gap: for small $k'_y$ close to the nodes of the bulk $d$-density-wave gap
the bound state wave function extends far into the system, while for larger
values of $k'_y$, corresponding to a larger bulk gap, 
the decay length is shorter. Furthermore we find surface bound states either 
for positive or negative $k'_y$, depending on the sign of the $d$-density-wave
amplitude at the surface.
A detailed analysis shows that a sizable part of the weight 
in the zero energy peak is
taken from relatively high energies $\approx 3 t$-$4t$. This appears to be 
comprehensible in view of the small extension of the bound state.

\subsection{Surface bound states, domain walls 
and domain wall bound states away from half filling}
\begin{figure}
\begin{center} 
\includegraphics[width=.49\textwidth]{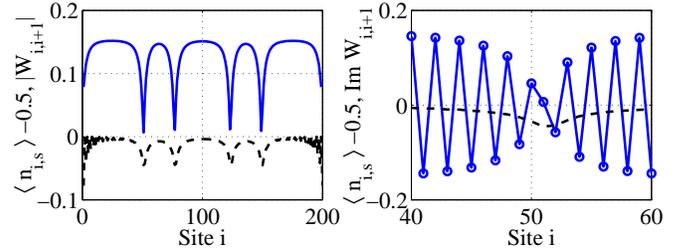}
\end{center} 
\caption{Left plot: Spatial variation of the  modulus of the 
$d$-density-wave amplitude $W_{i,i+1}$ (solid line) and the local
charge $\langle n_{i,s} 
\rangle$ (dashed line) per spin with respect to half-filling.) Right plot:
detailed structure of the domain wall showing the phase shift by $\pi$.
The data are calculated on a system with 200 sites in $x$-direction, $V=2t$, $\mu=-0.1t$ and $T=0.01t$.}
\label{mu10n200rs}
\end{figure}

Next we change the band filling away from half-filling by varying the
chemical potential $\mu$. Then the first striking
observation is that beyond a certain critical $\mu$
away from $\mu=0$ the self-consistent scheme converges to
non-homogeneous solutions where the $d$-density-wave pattern exhibits
domain walls  
with phase jumps of $\pi$, as shown in Fig.\ \ref{mu10n200rs} for $\mu=-0.1t$ 
corresponding to an average occupation number of $\approx 0.98$ particles per
site. Analyzing the charge distribution $\langle n_i \rangle$ (the
dashed lines in 
Fig.\ \ref{mu10n200rs} show the local deviation from half-filling), 
we find that the system chooses to
form locally half-filled regions with intact $d$-density-wave order. The
holes are concentrated in the domain walls. With increasing temperature, the
correlation length of the $d$-density-wave order gets longer and the domain
walls evaporate such that the bulk state becomes homogeneous again. 
If we move further away from half-filling, 
the domain wall density increases and finally, at another critical 
chemical potential (determined by the strength of the nearest neighbor
interaction 
$V$, for $V=2t$ at $\mu\approx -0.12t$), the local $d$-density-wave
order disappears completely.  

No attempt was undertaken to investigate the role of the boundary conditions
for the domain wall formation or to show that these
inhomogeneous solutions indeed correspond to global minima of the total free
energy. Further, if this was the case, the model should be extended to
include possible 
inhomogeneities in the direction along the surface as well. Here we do not
want to take this difficult route, 
or explore the mean-field phase diagram of this model in detail. 
Instead we will now assume that the domain walls along the [110] direction 
are stable or can be
stabilized by some additional interactions, and continue to analyze the local
excitation spectrum at the surface and in the domain wall.

\begin{figure}
\begin{center} 
\includegraphics[width=.48\textwidth]{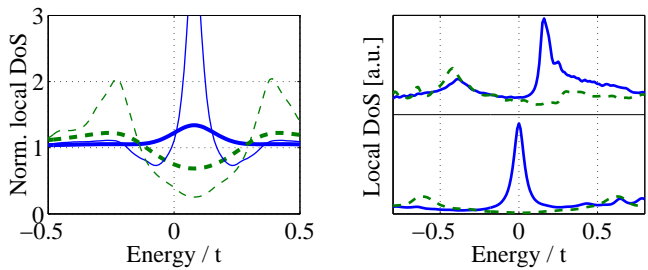}
\end{center} 
\caption{Left plot: Local density of states at the surface (solid lines) 
and in the bulk density of states (dashed lines) versus energy, for $V=2t$,
$\mu=-0.08t$ and $T=0.01t$ (thin lines) and $T=0.08t$ (thick lines),
normalized with respect to the values above the $d$-density wave transition at 
$T^{\mathrm{ddw}}_c \approx 0.125t$.
The local density of states was smeared out by
Lorentzians with width $2T$. Right plot, upper curves: surface (solid line) 
and  bulk (dashed line) density of states in the
$d$-density-wave state for $t'=-0.2t$ and
$\mu=-0.7t$ at $T=0.02t$. Right plot, lower curves: surface (solid line) and
bulk (dashed line) density of
states for homogeneous $d$-wave superconducting order 
coexisting with $d$-density-wave order 
(non-self-consistent) of comparable amplitude.
}
\label{mu8mu70dospl}
\end{figure}

According to the analytic calculation 
in Sec. II we expect the zero energy bound state to move to
finite energies. This is also the result of the self-consistent calculation, as 
shown in Fig.\ \ref{mu8mu70dospl} for $\mu=-0.08t$. For this chemical potential 
the bulk state is still homogeneous.
Apart from the bound state energy, the whole spectrum has moved and is 
now centered around a finite energy $\approx - \mu$, as also  
found by Zhu et al.\cite{zhu} and Wang\cite{wang} for the impurity case. For
the bulk system this shift is equivalent to the assumption of Chakravarty et
al. \cite{chakravarty} that upon hole-doping away from half-filling the 
$d$-density-wave dispersion stays rigid like in a semiconductor and that the
doped holes form Fermi surface pockets around the
nodal points of the half-filled state.

We now include a finite next nearest neighbor hopping $t'$. This has the
effect of destroying the $(\pi,\pi)$ nesting in the homogeneous state even at
half filling. Nevertheless, we find stable $d$-density-wave solutions,
in particular near the so-called van Hove filling for $\mu=4t'$ where the bulk
Fermi surface touches the van Hove singularities in the density of states at $(\pi,0)$ 
and $(0,\pi)$. For $V=2t$ and $t'=-0.2t$, stable solutions exist between
$\mu \approx -0.61t$ ($\langle n \rangle \approx 0.92$ per site) and $\mu
\approx -0.85t$ ($\langle n \rangle \approx 0.82$ per site). For band
fillings smaller or larger than these critical 
values the mean-field solutions for
 the $d$-density-wave state collapse rapidly with 
very narrow doping regions close to these critical dopings 
where inhomogeneous solutions occur. This suggests that the nonzero 
next nearest neighbor
hopping $t'$ inhibits the formation of well shaped domain walls with regions of
intact $d$-density wave order and charge accumulations in between.  
 In the right plot of Fig.\ \ref{mu8mu70dospl} we plot the
surface and bulk density of states for the homogeneous case 
$t'=-0.2t$ and $\mu=-0.7t$. Both curves 
have a large asymmetry with respect to zero energy and the peak in the
surface density of states is centered at comparably high excitation
energies. 

\begin{figure}
\begin{center} 
\includegraphics[width=.48\textwidth]{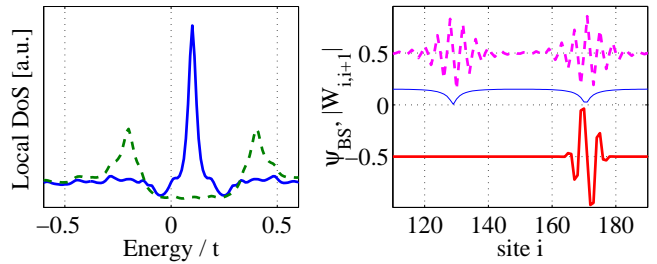}
\end{center} 
\caption{Left plot: Local density of states at a domain wall (solid line) 
and bulk density of states (dashed line) versus energy
The data are calculated on a system with 300 sites in $x$-direction, $V=2t$,
$\mu=-0.1t$ and $T=0.001t$. The local density of states was smeared out through
Lorentzians with width $0.01t$. Right plot: The upper ($k'_ya^*=-2.5$) and
lower ($k'_ya^*=-3$) curve (shifted by $\pm 0.5$) show
wave functions of states with different transverse wavevector $k$ 
contributing to the low energy peak at a domain wall. 
The thin line in the middle
shows the absolute magnitude of the $d$-density-wave amplitude $W_{i,i+1}$. 
}
\label{dwspec}
\end{figure}

Now let us turn back to the domain walls for $t'=0$ at sufficient doping away 
from half filling. As shown in Fig.\ \ref{dwspec}, the local density of states
at the domain wall exhibits a low energy peak similar to the one at the
surface. For a given transverse wavevector we can analyze the wave functions
contributing to this peak. Two of these wave functions for different transverse
wavevectors $k'_y$ are shown in the right plot of Fig.\ \ref{dwspec}: the lower 
curve belongs to a wave function with $k'_ya^*=-3$, while the upper curve is
for $k'_ya^*=-2.5$. 
The latter wave function has visible amplitude in two domain
walls. 
The difference between the two wave
functions can be easily understood: the wave function with larger transverse
momentum belongs to a larger bulk gap (in our coordinate system,
the bulk gap is maximal for $k'_ya^*=\pi$). Therefore the wave function
decays rapidly away from the domain wall. The 
wave function for the smaller $k'_y$ has a larger decay length and 
thus can effectively hybridize with solutions localized at the
neighboring domain wall. For even smaller values of the transverse wavevector, 
the wave functions corresponding to the sub-gap energy states 
distribute their weight over several domain walls.

\section{Conclusions}
We have calculated the local density of states in the presence of $d$-density
wave order on the two-dimensional square lattice. At [110]
surfaces we find zero energy (for perfect $(\pi,\pi)$ nesting of the Fermi
surface at half-filling) or low energy (away from perfect nesting) surface
states, which give rise to peaks in 
the calculated surface density of states, 
in good qualitative agreement with recent results for
the local density of states around impurities\cite{zhu,wang,morr}. 
The surface density of states is measurable in in-plane tunneling
experiments, as demonstrated for the high-$T_c$ cuprates 
in the superconducting 
state\cite{greene,alff,deutscher,yeh}. Similar to the proposals by Zhu et al.\cite{zhu}, 
Wang\cite{wang}
and Morr\cite{morr}, this provides another experimental test for possible
$d$-density-wave long range order 
in the pseudogap phase of the underdoped cuprates, 
as proposed by Chakravarty et al.\cite{chakravarty}. Since surface
effects crucially depend on the quality of the surface, it could still be that
the bound state features are wiped out by surface imperfections, even if
$d$-density-wave order is present. Moreover the temperature smearing reduces 
the low bias peaks considerably. Therefore we propose to perform
experiments on strongly underdoped samples with large pseudogap and 
low superconducting transition 
temperatures which have [110] surfaces that exhibit a zero-bias peak in the
superconducting state. The effect of surface imperfections on the latter
feature should be 
comparable to that on the $d$-density bound states. Thus,
if the superconducting state does exhibit an Andreev bound state one may expect
a $d$-density surface state in the pseudogap phase as well. We note that so
far the experimental results do not point towards $d$-density wave order:
published conductance measurements on underdoped YBCO films and LSCO single
crystals\cite{deutscher} and tunneling data on 001 tilt grain boundaries in 
underdoped YBCO\cite{alff} 
show zero energy peaks below the superconducting
$T_c$ and no feature above $T_c$.

For finite next-nearest neighbor hopping, the
peak in the calculated local density of states due to the surface states moves 
to rather large excitation energies, such that one might wonder if this peak
could still be distinguished from a gap edge similar to the one in the bulk
density of states. We think that the distinction is  still possible, 
because the peak due to the surface states has to move to zero energy when the system enters
the $d$-wave superconducting state. In contrast with that the gap edge feature 
due to the extended bulk states should move to higher excitation energies 
in the superconducting state (see lower curves in the right plot of 
Fig.\ \ref{mu8mu70dospl}), 
if  the latter is indeed formed on top of the $d$-density-wave order, 
as suggested 
by Chakravarty et al.\cite{chakravarty}. One may even try
to observe a shift of the bound state feature -- if existent above the
superconducting $T_c$ --
 from nonzero energies due to the imperfect nesting 
to zero energy when one goes from the $d$-density-wave phase into the $d$-wave
superconducting state.
Another interesting 
experimental question might be the search for domain walls of the 
$d$-density-wave
order parameter along the [110] direction which again should give rise to low
energy peaks in the local $c$-axis tunneling density of states. 
We note however that our calculation is not a
stringent proof of the existence of domain walls for the 
reasons explained above. 
\\[3mm]
It is a pleasure to thank Chetan Nayak,  Ying Liu, Masatsura Igami, Manfred
Salmhofer and
Katsunori Wakabayashi for helpful
discussions. This work was financially supported
 by the Swiss National Science Foundation.

\end{document}